\documentclass[journal]{IEEEtran}

\ifCLASSINFOpdf
\else
\fi

\usepackage{graphicx}

\usepackage{longtable}
\usepackage{subfigure}
\usepackage{url}
\usepackage{booktabs}
\usepackage{tabularx}
\usepackage{amsthm}
\usepackage{amsmath}
\usepackage{cite}

\usepackage{lineno,hyperref}
\usepackage{enumerate}

\newtheorem{definition}{Definition}[section]
\newtheorem{example}{Example}[section]

\usepackage[]{changes}
  \definechangesauthor[name={Tao Wen}, color=orange]{add}

\usepackage{soul}
  \soulregister\cite7
  \soulregister\ref7

\begin{document}
\title{Evaluating the Vulnerability of Communities in Social Networks by Gravity Model}

\author{Tao Wen
\thanks{Manuscript received November XX, 2019; revised XXX XX, 20XX. \emph{(Corresponding Author: Tao Wen)}}
\thanks{T. Wen is with Ministry of Education Key Laboratory for Intelligent Networks and Network Security, Xi’an Jiaotong University, Xi'an, 710049, China. (E-mail: taowen@stu.xjtu.edu.cn.)}
}

\markboth{Journal of \LaTeX\ Class Files,~Vol.~14, No.~8, August~2015}
{Shell \MakeLowercase{\textit{et al.}}: Bare Demo of IEEEtran.cls for IEEE Journals}

\maketitle
\begin{abstract}
With the development of network science, the various properties of complex networks have recently received extensive attention. Among these properties, the vulnerability of the communities (VoCs) is particularly important. In the conventional research, only parts of structural features of the community rather than multiple aspects are considered in the evaluating model. However, in reality, the impact on the VoC is multifaceted, not only its own structure property, but also the influence of other communities. In order to better model the influence between communities, so as to evaluate the VoCs in the social network, a gravity-based community vulnerability evaluation model is proposed in this paper. In this proposed model, three different aspects of the factor are considered, i.e. the number of edges inside the community, the number of edges connected neighboring communities, and the gravity index (GI) of each community, which correspond to the interior information, small scale interaction relationship, and large scale interaction relationship of communities. By means of the Jensen-Shannon divergence (JSD) and log-sigmoid transition (LST) function, the abstract distance (AD) between each pair of communities can be calculated to construct the community network (CN). With the usage of gravity model, the GI of each community which describes the large scale interaction relationship can be obtained. Eventually, the community vulnerability degree and order can be calculated by this proposed model, and the sensitivity of weighting parameters is analyzed by Sobol' indices. In particular, this proposed method can degenerate to the classical method with the setting of weighting parameters. The effectiveness and reasonability of this proposed model are demonstrated by several real world complex networks. 
\end{abstract}

\begin{IEEEkeywords}
  Complex network, Community vulnerability, Gravity model, Jensen-Shannon divergence, Log-sigmoid transition.
\end{IEEEkeywords}

\IEEEpeerreviewmaketitle

\section{Introduction}\label{Sec_introduction}


With the development of the Internet, socialization has become an indispensable part of people's daily life, such as sharing news on Twitter, chatting with friends via Telegram, and so on \cite{Zhao2018Real}. In order to quantify people's performance in social, social networks are immediately proposed to describe individuals' relationships. In social networks, each individual has a closely related group, and it is called community. The community structure is widespread in the network, and it has become a popular topic in the field of network science \cite{Boccaletti2006Structure,Watts1998Collective,Jiang2015Diffusion}. In general, the research in the network community is mainly divided into two fundamental issues. The first issue is how to detect the community structure in the network effectively and accurately, that is, how to divide the network into high-density groups with different sizes. Numerous algorithms have been developed to address this problem in different types of networks. More specifically, Newman \cite{Newman2006Modularity} developed a series of models to detect community structure which is the basis of this field. Liu \emph{et al.} \cite{Liu2014Multiobjective} proposed a multi-objective evolutionary algorithm to detect community structure in social networks which can deal with the positive and negative links in overlapping communities. Fortunato \cite{Fortunato2010Community} summarized most interdisciplinary methods to detect community structure from the definition of the main elements, and applied these algorithms in lots of real world networks. There are still lots of detection algorithms based on machine learning, such as Z-network model \cite{Jiang2019Znetwork}, latent space graph regularization \cite{Yang2015Unified}, and particle swarm optimization \cite{Zeng2019Consensus}. The other issue is how to quantitatively describe the common properties in each community, because these community properties will affect the judgment of the whole network structure. For example, the similarly of each pair of nodes has been measured to find the same group across multi-layer social network \cite{Liu2018We,kang2019Znumbers,wentao2019similar}. The subset of high-propagation capabilities individuals has been identified to reduce the scale of disaster and rumour transmission \cite{Chen2019Ant,wen2019Identification,Zhang2018Optimizing}. The resilience of community has been quantified to describe how the initial set of community structures survived and returned after the disruption \cite{Rocco2018Quantifying,Zhang2018uncertainty,De2018Estimating}. The key opinion leaders have been identified to draw specific types of audiences, which can form a long-term relationship and govern the success or failure of social activities \cite{De2015Trust,Liu2018Identifying,Wang2019Finding}.


Among these properties of community, the vulnerability of community has gradually attracted researchers' attention, because this property can visualize the important components of social networks. By evaluating the vulnerability of each community, resources of the network can be prioritized which can reduce the damage for the network, and the performance of each component in networks can be well protected or improved. Thus, lots of researches have been focused on evaluating community vulnerability and corresponding protecting methods \cite{Rocco2011Assessing,wentao2018evaluating,Boccaletti2007Multiscale}. Recently, Che \emph{et al.} \cite{Yanbo2019Vulnerability} gave a nondimensionalized scoring standard based on Analytic Hierarchy Process and Artificial Neural Network to assess the vulnerability of urban power grid, which is in the field of machine learning. Then, Rocco \emph{et al.} \cite{Rocco2011Vulnerability} developed a qualitative metric to evaluate the community vulnerability which focuses on the connectivity between every two connected communities. This qualitative metric provides some insight into the relative strength of each community, and has been used in several real world complex networks. Thus, connectivity has been shown to be an important consideration in individuals' resistance to disasters in the network, i.e. the vulnerability of community. Based on this qualitative metric, Wei \emph{et al.} \cite{Wei2018Measuring} proposed a generalized model which considers both the outer connectivity of a community and its inner structure. This model combines 5 factors (3 internal factors and 2 external factors) into 1 comprehensive metric to evaluate the community vulnerability which is suitable not only for unweighted networks but also for weighted networks. 


In order to better describe the relationship between communities, so as to evaluate the community vulnerability, a novel method which is called gravity-based community vulnerability evaluation (GBCVE) model is developed in this paper, which involves from the community detection to the final vulnerability evaluating and ranking of each community. Specifically, the community structure is detected by Newman's modularity \cite{Newman2004Fast} from the social network. The basic information can be obtained from the structure directly, which contains the number of edges inside the community (EIC) and the number of edges outside the community (EOC). Then, the abstract distance (AD) which describes the difference between every two communities is obtained by Jensen-Shannon divergence (JSD) and log-sigmoid transition (LST) technique. Thus, the community network (CN) which is a fully-connected undirected network can be constructed by the AD. After that, the gravity index (GI) of each community is calculated based on the gravity model from the CN. This proposed model would combine these three factors (EIC, EOC, GI) with weighting parameters, which can be used in different situations. This proposed method would degenerate to the classical method under special weighting parameter setting. The EIC, EOC, and GI correspond to different aspects of consideration information, i.e. the interior information of each community, the small scale interaction relationship, and large scale interaction relationship respectively. Eventually, the vulnerability of each community can be calculated by this proposed model, and the vulnerability order can be obtained by fuzzy ranking algorithm, which is more reasonable to describe their vulnerability relationship between different communities. In addition, the sensitivity of these weighting parameters is analyzed by Sobol' indices. After two real world complex networks evaluation, the reasonability and effectiveness of this proposed model can be demonstrated. The results show the superiority of this proposed method, and it is an effective model to measure the property of communities.


The rest of this paper is organized as follows. The definition of this problem is given in Section \ref{Sec_problem}. Section \ref{Sec_methodology} develops this community vulnerability evaluation model in detail, which is divided into five parts. The experimental evaluations which contain two real world complex networks are given in Section \ref{Sec_evaluations}, and the results show the effectiveness and reasonability of this proposed model. The conclusions and future work are discussed in Section \ref{Sec_conclusion}.



\section{Problem Definition}\label{Sec_problem}

\begin{figure}[!htbp]
  \centering
  \includegraphics[width=0.3\textwidth]{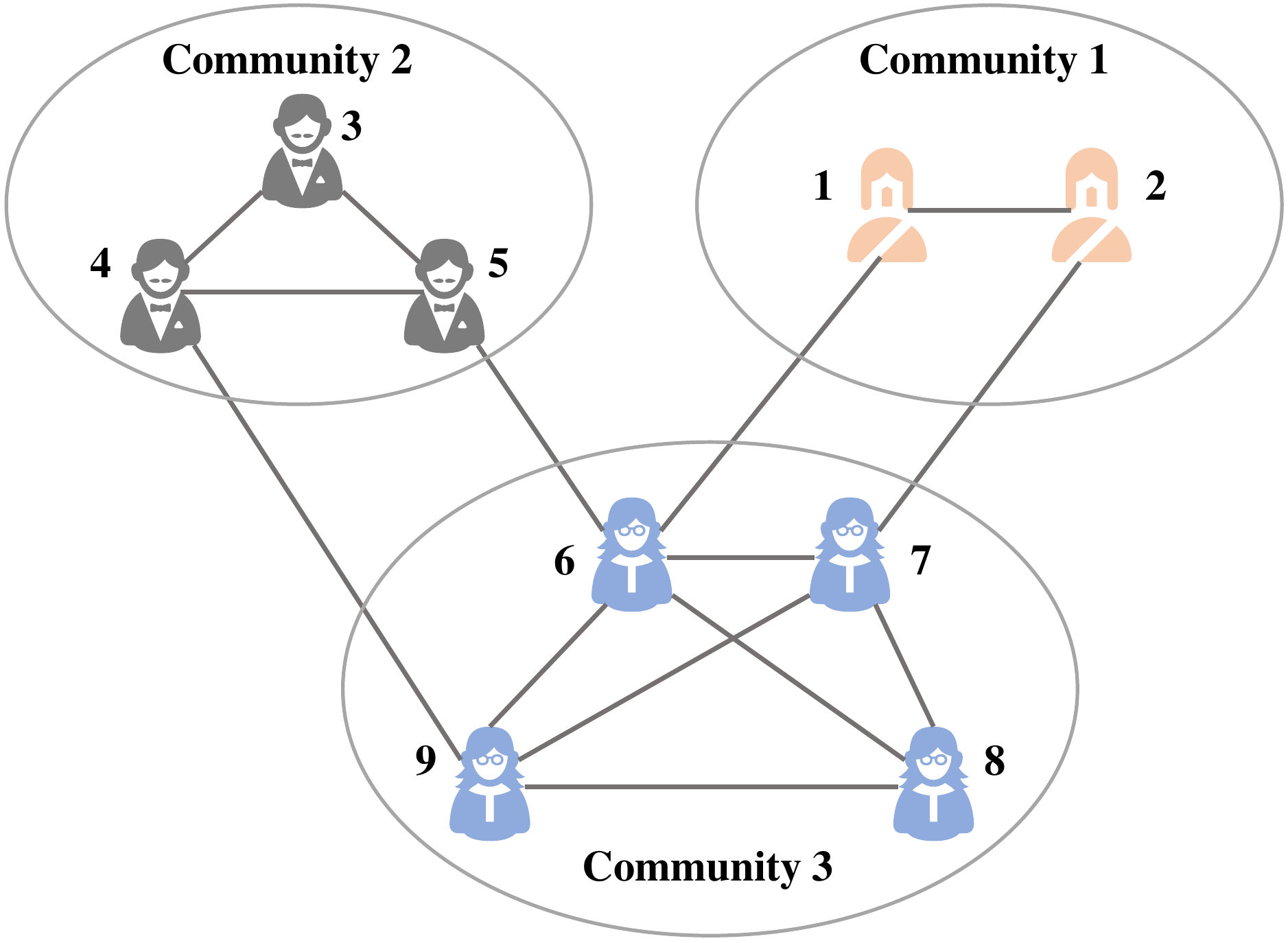}\\
  \caption{\textbf{The example social network with 9 nodes.}}
  \label{Fig_example_network}
\end{figure}

For a given social network $G(N,E)$, it is composed of multiple social individuals and the relationships (links) between individuals. Each individual in the network has different connection strengths, so the network can be divided into several communities. One example social network with 9 individuals and 14 links is shown in Fig. \ref{Fig_example_network}, and this social network can be divided into 3 communities (${c_1}$, ${c_2}$, and ${c_3}$) by Newman's modularity method \cite{Newman2004Fast}. The purpose of this paper is to evaluate the \emph{vulnerability of communities} (VoCs) by considering multiple structure properties of each community, and get the vulnerability order in social networks which can guide the establishment of social networks.

According to the above introduction and definition, the problem is defined as follows:

\emph{Input:} The input information is the structure of social network $G(N,E)$, where $N$ is the set of individuals, and $E$ is the set of relationships between individuals. The topological relationship of the social network is given by the adjacency matrix $A$, and the element ${a_{ij}}$ in $A$ indicates whether there is a connection between individual $i$ and individual $j$.

\emph{Output:} When the social network is divided in several communities, this proposed method can evaluate the \emph{vulnerability of community ${c_i}$} ($Vo{C_i}$), and give the vulnerability order in the network.

\section{Methodology}\label{Sec_methodology}

The community structure of social networks is key to be considered. In the real world, every individual has their own community, and the VoCs also affects individuals' activities. Thus, a novel method called \emph{gravity-based community vulnerability evaluation} (GBCVE) model is proposed in this paper to consider: \emph{1) the property of community itself; 2) the relationships between the chosen community and other communities (large scale and small scale).} Adequate consideration of community information makes GBCVE more reasonable and effective to obtain the vulnerability results. 

For a given social network $G(N,V)$, the VoCs can be obtained by the following steps:

\begin{enumerate}[1)]
  \item \textbf{Divide the network into several communities}. \\
  The whole network $G$ can be divided into multiple non-intersecting communities, and each individual belongs to only one community ${c_i}$.
  \item \textbf{Obtain the basic information of each community}.\\
  With the obtained community structure, the basic information of community can be obtained, which contains the \emph{number of edges inside the community} (EIC), and the \emph{number of edges outside the community} (EOC).
  \item \textbf{Construct community network by the abstract distance}.\\
  According to the structure of communities, the \emph{Jensen-Shannon divergence} (JSD) between communities can be obtained by Eq. (\ref{Eq_JSD}) to describe the difference between communities, then, the JSD can be transformed to \emph{abstract distance} (AD) to construct \emph{community network} (CN) by \emph{log-sigmoid transition} (LST) function in Eq. (\ref{Eq_log})
  \item \textbf{Build the large scale relationship between different communities}.\\
  Given the computed AD between communities, the large scale relationship (\emph{gravity index}, GI) between each pair of communities can be built by Eq. (\ref{Eq_gravity_model}) based on the \emph{gravity model} (GM).
  \item \textbf{Compute the vulnerability of each community}.\\
  With the obtained three factors of each community (EIC, EOC, and GI), the VoC can be calculated based on the comprehensive consideration of the property of the community.
  \item \textbf{Obtain the fuzzy ranking of community vulnerability}.\\
  After the VoC calculation process, the fuzzy ranking order of community vulnerability can be obtained.
\end{enumerate}

\begin{figure}[!htbp]
  \centering
  \includegraphics[width=0.5\textwidth]{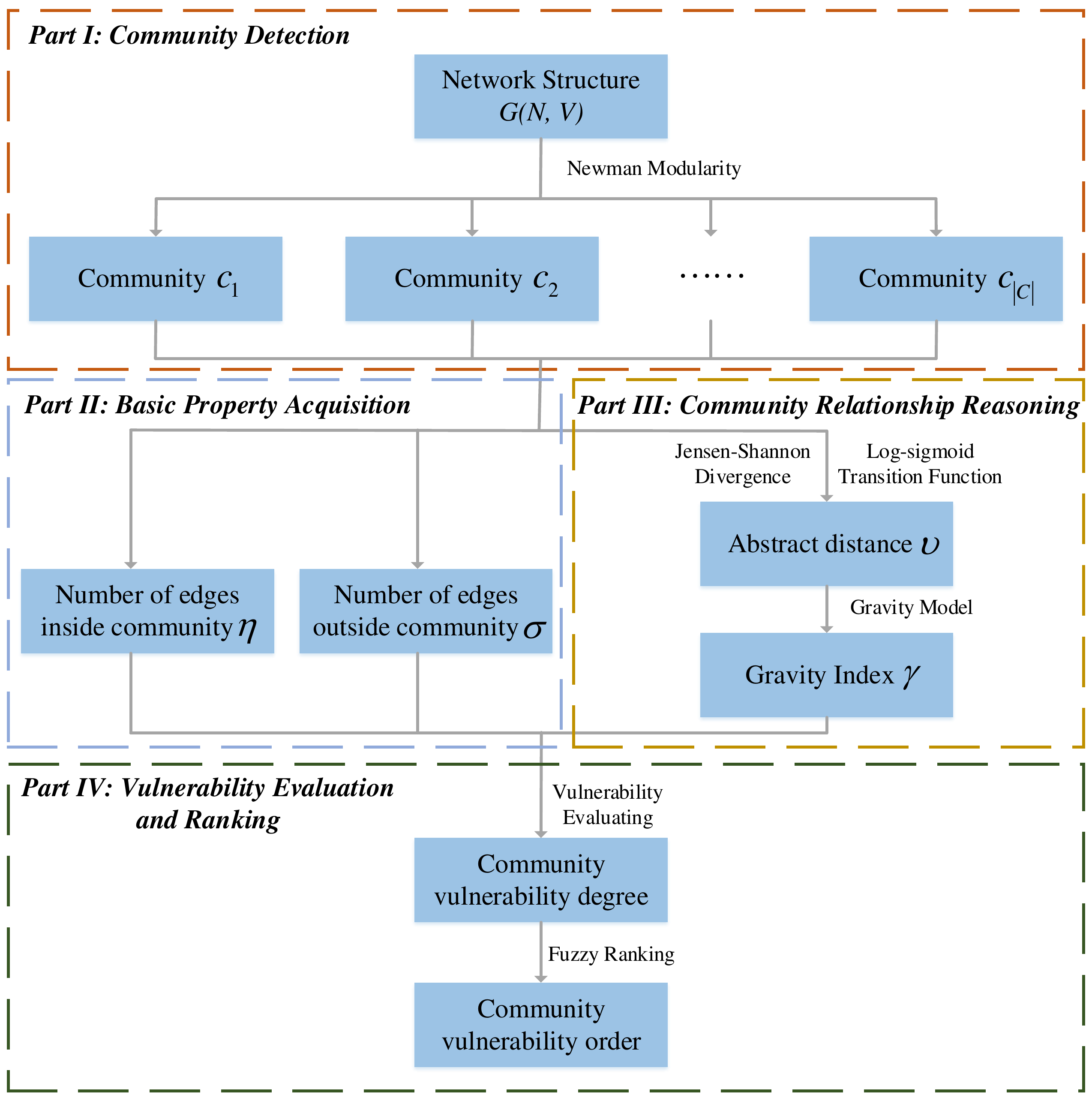}\\
  \caption{\textbf{The framework of this proposed GBCVE model.}}
  \label{Fig_framework}
\end{figure}

The framework of this proposed GBCVE model is shown in Fig. \ref{Fig_framework}. 

In order to introduce the model in detail, the rest of this section is divided into five main subsections from Section \ref{Sub_Method_Detecting} to \ref{Sub_Method_sensitive}. Specifically, Section \ref{Sub_Method_Detecting} introduces how to get the community structure from the social network, which corresponds \emph{Step 1}. Section \ref{Sub_Method_Basic} obtains the basic information of each community, which shows \emph{Step 2}. The community relationship can be measured by divergence, transition technique, and gravity model in Section \ref{Sub_Method_gravity}, which contains \emph{Step 3} and \emph{Step 4}. The vulnerability evaluating and ranking process are given in Section \ref{Sub_Method_vulnerability}, corresponding to \emph{Step 5} and \emph{Step 6}. Eventually, the sensitivity of this proposed model is analyzed in Section \ref{Sub_Method_sensitive}.


\subsection{Part I: Communities Detection} \label{Sub_Method_Detecting}

A social network can be denoted as $G(N,E)$, where $N = \{ 1,2, \cdots ,\left| N \right|\}$ and $V = \{ 1,2, \cdots ,\left| V \right|\}$ are the set of nodes and edges respectively, $\left| N \right|$ and $\left| V \right|$ are the whole number of nodes and edges respectively. The topological relationship in the network is represented by the adjacency matrix $A$, whose size is $\left| N \right| \times \left| N \right|$. The element ${a_{ij}}$ shows whether there is a edge between node $i$ and node $j$, i.e. ${a_{ij}} = 1$ means the existence of edges and vice versa.

\begin{definition} \label{Def_modularity}
  The modularity of network with $\left| C \right|$ communities is denoted as $Q$ \cite{Newman2004Fast}, and is defined below:
  \begin{equation} \label{Eq_modularity}
    Q = \sum\limits_{{c_k} = 1}^{\left| C \right|} {\left( {\frac{{\left| {{E_{{c_k}}}} \right|}}{{\left| E \right|}} - {{\left( {\frac{{\sum\limits_{i \in {N_{{c_k}}}} {{d_i}} }}{{2\left| E \right|}}} \right)}^2}} \right)} 
  \end{equation}
  where ${{c_k}}$ is the chosen community, ${\left| C \right|}$ and ${\left| E \right|}$ is the total number of communities and edges in the network respectively, ${\left| {{E_{{c_k}}}} \right|}$ is the number of edges inside the community ${c_k}$, ${{N_{{c_k}}}}$ is the set of nodes in community ${c_k}$, and ${d_i}$ is the degree of node $i$. The degree of node means the number of edges connected with chosen node.
\end{definition}

The value of $Q$ can be used as a good indicator to measure the existence of the community structure, and different values of $Q$ show whether the existing community division is reasonable. 

\begin{itemize}
  \item $Q = 0$ represents all nodes (i.e. $\left| N \right|$ nodes) are divided into one single community, and there will be no community structure in the social network.
  \item $Q \in (0,1)$ means the existence of community structure, and different values of $Q$ indicate the strength of the community structure.
  \item $Q = 1$ indicates the existing community structure is very strong.
\end{itemize}

In addition, Girvan and Newman \cite{Newman2004Finding} gave the suggestion that the value of $Q$ between 0.2 and 0.7 indicates a good community structure in the social network. The algorithm to detect the community structure is to find the change of $Q$ value, and the detail steps are shown below:

\begin{enumerate}[Step 1:]
    \item Treat every point in the network as a separate community, i.e. there is $\left| N \right|$ communities at the beginning, and obtain ${Q_0}$ in this situation.
    \item Combine any two communities into a community in turn, and calculate ${Q_t}(i,j)$ of the integrated community structure (combined community $i$ and $j$ in \emph{t}th integration).
    \item Obtain the modularity change $\Delta {Q_t}(i,j) = {Q_t}(i,j) - {Q_{t - 1}}$ before and after the integration from the community structure.
    \item Find the maximum value of $\Delta {Q_t}$ in the \emph{t}th integration, i.e. $\Delta {Q_t} = \mathop {\max }\limits_{i,j} \Delta {Q_t}(i,j)$, and this structure with modularity ${Q_t}$ is chosen as the integration method.
    \item Repeat Step 2 to Step 4 until $\Delta {Q_t} < 0$.
\end{enumerate}

\begin{example} \label{Example_Q}
  The social network in Fig. \ref{Fig_example_network} is used as an example to illustrate the community detection algorithm. The integration process and the change of $Q$ is shown in Table \ref{table_integration}. The community structure in each time is the existing community division, and these communities shown in \textbf{bold} form will integrate at next time. ${Q_t}$ is calculated by the community structure at this time. Take time at 6 as an example, the process to obtain ${Q_t}$ is:
  \begin{equation*} \label{Eq_example_Q_t}
    \begin{split}
      \displaystyle {Q_6} & = \left( {\frac{1}{{14}} - {{\left( {\frac{4}{{28}}} \right)}^2}} \right) + \left( {\frac{3}{{14}} - {{\left( {\frac{8}{{28}}} \right)}^2}} \right)\\
      \displaystyle & + \left( {\frac{6}{{14}} - {{\left( {\frac{{16}}{{28}}} \right)}^2}} \right) = 0.2857
      \end{split}
  \end{equation*}
  and the modularity change $\Delta {Q_6} = {Q_6} - {Q_5} = 0.2857 - 0.2117 = 0.0740$. When the modularity change $\Delta {Q_t} < 0$, the integration process will stop, and the final community structure will appear. Thus, the social network is finally divided into three communities, and they are \{1, 2\}, \{3, 4, 5\}, \{6, 7, 8, 9\} respectively.

\begin{table}[!htbp]
  \centering
  \caption{\textbf{The integration process and the change of $Q$.}}
  \resizebox{0.5\textwidth}{!}{
  \begin{tabular}{cccc}
  \hline
  \hline
  Times \emph{t} & Communities structure & ${Q_t}$ &  $\Delta {Q_t}$ \\
  \hline
  0      & \textbf{\{1\}}, \textbf{\{2\}}, \{3\}, \{4\}, \{5\}, \{6\}, \{7\}, \{8\}, \{9\} & -0.1224 & -- \\
  1      &  \{1, 2\}, \textbf{\{3\}}, \textbf{\{4\}}, \{5\}, \{6\}, \{7\}, \{8\}, \{9\} & -0.0612 & 0.0612 \\
  2      &  \{1, 2\}, \textbf{\{3, 4\}}, \textbf{\{5\}}, \{6\}, \{7\}, \{8\}, \{9\} & -0.0051 & 0.0561 \\
  3      &  \{1, 2\}, \{3, 4, 5\}, \{6\}, \textbf{\{7\}}, \textbf{\{8\}}, \{9\} & 0.0995 & 0.1046 \\
  4      &  \{1, 2\}, \{3, 4, 5\}, \{6\}, \textbf{\{7, 8\}}, \textbf{\{9\}} & 0.1403 & 0.0408 \\
  5      &  \{1, 2\}, \{3, 4, 5\}, \textbf{\{6\}}, \textbf{\{7, 8, 9\}} & 0.2117 & 0.0714 \\
  6      &  \{1, 2\}, \textbf{\{3, 4, 5\}}, \textbf{\{6, 7, 8, 9\}} & 0.2857 & 0.0740 \\
  7      &  \{1, 2\}, \{3, 4, 5, 6, 7, 8, 9\} & 0.2653 & -0.0204 \\
  \hline
  \hline
  \end{tabular}}
  \label{table_integration}
\end{table}
\end{example}

\subsection{Part II: Basic Property Acquisition} \label{Sub_Method_Basic}

After the community structure of social network is obtained, the basic information of each community can be obtained to describe their properties. The \emph{number of edges inside the community} (EIC) and the \emph{number of edges outside the community} (EOC) are two significant properties in different aspects to describe community vulnerability. 

\begin{definition} \label{Def_EIC}
  The EIC for community ${c_i}$ is obtained as below:
  \begin{equation} \label{Eq_EIC}
    {\eta _i} = \sum\limits_{j,k \in {c_i}} {{a_{jk}}} 
  \end{equation}
  where ${{a_{jk}}} $ represents whether there is an edge between node $j$ and node $k$, and node $j$, $k$ belong to community ${c_i}$. This means the edge is completely inside the community, not connected to the outside, indicating the connectivity within the community. Thus, EIC demonstrates the interior information of community.
\end{definition}

\begin{definition} \label{Def_EOC}
  The EOC for community ${c_i}$ is obtained as below:
  \begin{equation} \label{Eq_EOC}
    {\sigma _i} = \sum\limits_{j \in {c_i},k \notin {c_i}} {{a_{jk}}} 
  \end{equation}
  which is different with the EIC, the EOC describes the accessibility between the chosen community and other communities (node $j$ belongs to community ${c_i}$, but node $k$ belongs to other communities except community ${c_i}$). Large ${\sigma _i}$ means this community ${c_i}$ is convenient to communicate with other communities. Thus, EOC demonstrates the small scale interaction relationship of community. 
\end{definition}

\begin{example} 
  For the social network in Fig. \ref{Fig_example_network}, it has been divided into three communities in Example \ref{Example_Q}. ${\eta _i}$ and ${\sigma _i}$ of these three communities can be obtained by the community structure, and they are $\eta  = [1,3,6]$ and $\sigma  = [2,2,4]$ respectively.
\end{example}

\subsection{Part III: Large Scale Community Relationship Reasoning} \label{Sub_Method_gravity}

After obtaining the information within the community (EIC) and with neighboring societies (EOC), the relationship between each pair of communities, i.e. the large scale information, is also significant to evaluate each community vulnerability. To measure the large scale relationship, JSD, LST, and GM are applied in this section. The details are shown below:

\subsubsection{Community Abstract Distance}

The distance between communities can be regarded as the physical distance, i.e. the sum of the length of the edges in the network, but it only reflects the distance between communities in space. In this section, community properties are considered to calculate the abstract distance (AD) between communities. 

Assume that in all communities, the community with the largest number of nodes has $\kappa $ nodes, i.e. $\kappa  = \mathop {\max }\limits_{{c_k}} \left| {{N_{{c_k}}}} \right|$. There is $\left| {{N_{{c_i}}}} \right|$ nodes in community ${c_i}$, and the probability set for community ${c_i}$ is shown below:
\begin{equation} \label{Eq_probability_set}
  {P_i} = [{p_i}(1),{p_i}(2), \cdots ,{p_i}(m), \cdots ,{p_i}(\kappa ),]
\end{equation}
This setting makes all probability sets the same size, i.e. $\kappa$ elements. $\left| {{N_{{c_i}}}} \right|$ is usually less than $\kappa $, and equals $\kappa $ when ${c_i}$ is the largest community. Thus, some element of ${P_i}$ would equal to zero when $\left| {{N_{{c_i}}}} \right| < \kappa$. The detail definition of the element of ${P_i}$ is given below:
\begin{equation} \label{Eq_element_P_i}
  {p_i}(t) = \left\{ {\begin{array}{*{20}{c}}
    {\frac{{{d_t}}}{{\sum\limits_{t \in {N_{{c_i}}}} {{d_t}} }}}&{t \le \left| {{N_{{c_i}}}} \right|}\\
    0&{t > \left| {{N_{{c_i}}}} \right|}
    \end{array}} \right.
\end{equation}
where ${d_t}$ is the degree of node $t$. More detail, each node in community has one probability value, and the rest of elements equal to zero to complete the probability set. Then, the probability set ${P_i}$ should be sorted in descending order, and it is denoted as ${P_i}^\prime $, which can be shown below:
\begin{equation} \label{Eq_descending_P}
  {P_i}^\prime  = [{p_i}^\prime (1),{p_i}^\prime (2), \cdots ,{p_i}^\prime (t), \cdots ,{p_i}^\prime (\kappa ),]
\end{equation}
This is because the order of probability set would affect the JSD between two probability set, and the descending order can eliminate the error caused by the sequence.

\begin{definition} \label{Def_JSD}
  The Jensen-Shannon divergence (JSD) \cite{Endres2003new} can measure the difference between probability distribution in information theory. In this paper, JSD is applied to measure the difference between two community structure, which is denoted as ${\mu _{ij}} $ and defined as follows:
  \begin{equation} \label{Eq_JSD}
    \begin{split}
      \displaystyle    {\mu _{ij}}  & = {D_{JS}}({P_i}^\prime ||{P_j}^\prime )\\
      \displaystyle      & = \frac{1}{2} \times \left( {\sum\limits_{t = 1}^{\kappa '} {{p_i}^\prime (t)\ln \frac{{{p_i}^\prime (t)}}{{{p_{con}}^\prime (t)}}}  + \sum\limits_{t = 1}^{\kappa '} {{p_j}^\prime (t)\ln \frac{{{p_j}^\prime (t)}}{{{p_{con}}^\prime (t)}}} } \right)
      \end{split}
  \end{equation}
  where ${P_{con}}^\prime $ is obtained as follows:
  \begin{equation}
    {P_{con}}^\prime  = \frac{{{P_i}^\prime  + {P_j}^\prime }}{2}
  \end{equation}
  ${P_i}^\prime $ and ${P_j}^\prime $ are two descending order probability sets for ${c_i}$ and ${c_j}$, ${{p_i}^\prime (t)}$ and ${{p_j}^\prime (t)}$ are the elements in the probability set which are obtained from Eq. (\ref{Eq_element_P_i}). ${\kappa '}$ is the smaller one between $\left| {{N_{{c_i}}}} \right|$ and $\left| {{N_{{c_j}}}} \right|$, i.e. $\kappa ' = \min \{ \left| {{N_{{c_i}}}} \right|,\left| {{N_{{c_j}}}} \right|\} $. The chosen of ${\kappa '}$ is to avoid ${p_i}^\prime (t)/{p_{con}}^\prime (t)$ or ${p_j}^\prime (t)/{p_{con}}^\prime (t)$ being $0$ or infinity, thus outputting out-of-range logarithmic values. 
\end{definition}

Compared with relative entropy (Kullback-Leibler divergence, KLD), the JSD has the following difference:
\begin{itemize}
  \item JSD is symmetrical, unlike KLD is asymmetrical, i.e. ${D_{JS}}({P_i}^\prime ||{P_j}^\prime ) = {D_{JS}}({P_j}^\prime ||{P_i}^\prime )$.
  \item The range of JSD is $[0, 1]$, when two probability distributions are the same, ${D_{JS}}$ equals to 0, and ${D_{JS}} = 1$ means these two probability distributions are completely different.
\end{itemize}

With the obtained JSD, log-sigmoid transition (LST) function which is also called S-type growth curve is used as a transition function for the distance between communities. Because of the monotonically increasing character and inverse function, LST is wildly in information science.

\begin{definition} \label{Def_log}
  The AD ${\upsilon _{ij}}$ obtained by LST function is defined as follows:
  \begin{equation} \label{Eq_log}
    {\upsilon _{ij}} = \frac{1}{{1 + {e^{ - \varphi  \cdot {\mu _{ij}}}}}}, i \ne j
  \end{equation}
  where ${{\mu _{ij}}}$ is the JSD between communities, and $\varphi $ is the fitting parameter to adjust the performance of this model. $i \ne j$ means LST function would not change the AD of community itself. With the change of $\varphi $, the AD between communities will also change. The fitting parameter $\varphi $ is chosen as 3 in this model to achieve a better effect. The figure corresponding to the function with $\varphi = 3$ is shown in Fig. \ref{Fig_log}. 
  \begin{figure}[!htbp]
    \centering
    \includegraphics[width=0.45\textwidth]{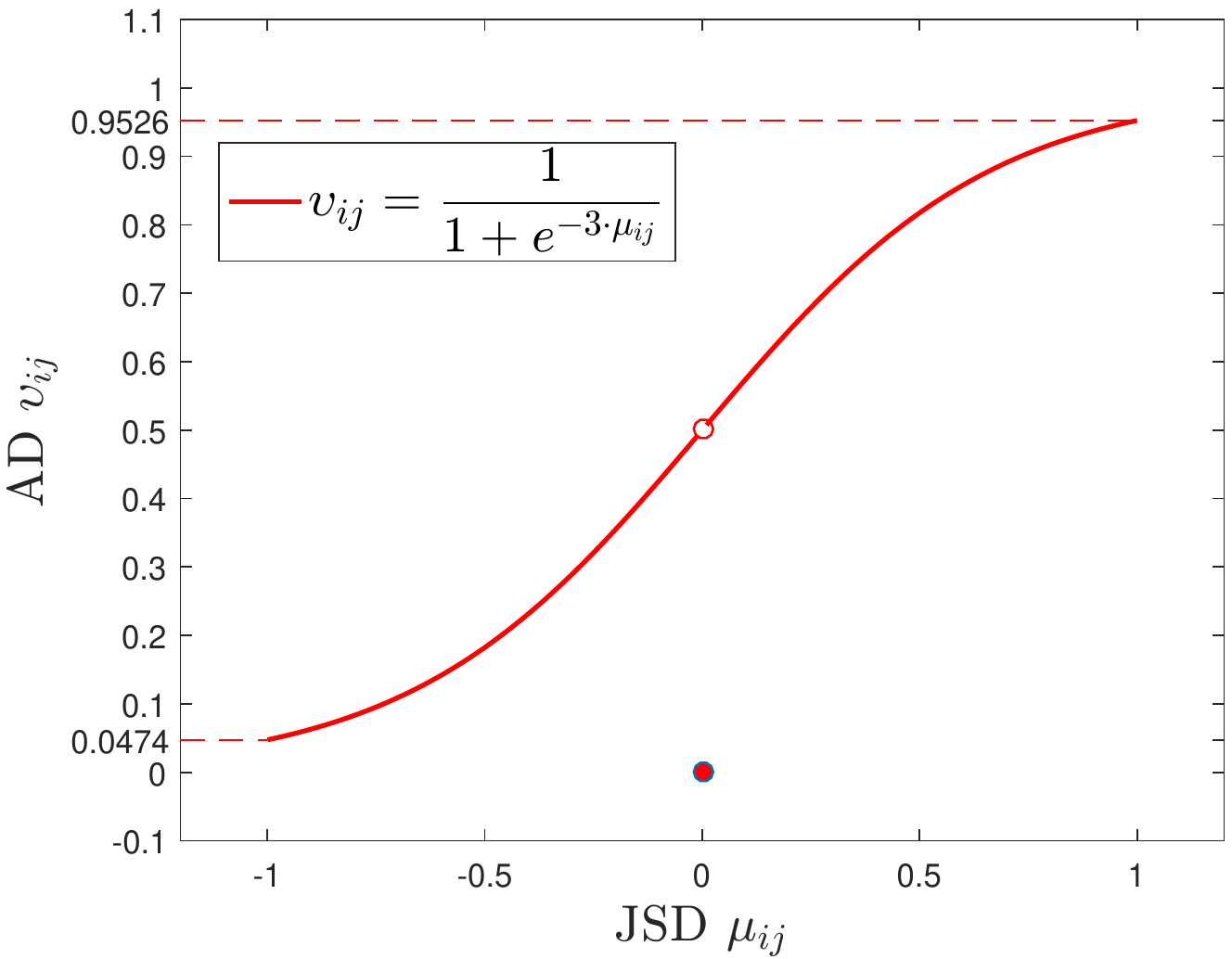}\\
    \caption{\textbf{The log-sigmoid function expression with $\varphi = 3$.}}
    \label{Fig_log}
  \end{figure}

\end{definition}

Thus, the AD between any two communities can be obtained to construct \emph{community network} (CN), and the distance to describe the difference of communities can be shown by ${\upsilon _{ij}}$. Here one example is given to illustrate the algorithm.

\begin{example}
  With the obtained community structure from Example \ref{Example_Q} for the social network shown in Fig. \ref{Fig_example_network}, ${c_2}$ and ${c_3}$ are chosen as example. The descending order of probability set for ${c_2}$ and ${c_3}$ are ${P_2}^\prime  = [\frac{1}{3},\frac{1}{3},\frac{1}{3},0]$ and ${P_3}^\prime  = [\frac{1}{4},\frac{1}{4},\frac{1}{4},\frac{1}{4}]$ respectively. Then, ${P_{con}}^\prime $ can be calculated as ${P_{con}}^\prime  = [\frac{7}{{24}},\frac{7}{{24}},\frac{7}{{24}},\frac{1}{8}]$.

  Because $\left| {{N_{{c_2}}}} \right|$ and $\left| {{N_{{c_3}}}} \right|$ equal to 3 and 4 respectively, ${\kappa '}$ would be assigned 3. Then the JSD ${\mu _{23}}$ between ${c_2}$ and ${c_3}$ is obtained below:
  \begin{equation*}
    \begin{split}
      \displaystyle {\mu _{23}} = {\mu _{32}} & = \frac{1}{2} \times (3 \times \frac{1}{3} \times \ln \frac{{1/3}}{{7/24}} + 3 \times \frac{1}{4} \times \ln \frac{{1/4}}{{7/24}})\\
      \displaystyle & = 0.0090
      \end{split}
  \end{equation*}
  The JSD ${\mu _{23}}$ would be transformed to AD ${\upsilon _{23}}$ with LST function:
  \begin{equation*}
    {\upsilon _{23}} = \frac{1}{{1 + {e^{ - 3 \times 0.0090}}}} = {\rm{0}}{\rm{.5067}}
  \end{equation*}
  and the AD matrix for this community structure is obtained as follows:
    \begin{equation*}
      \displaystyle \upsilon  = \left[ {\begin{array}{*{20}{c}}
        0&{0.5126}&{0.5318}\\
        {0.5126}&0&{0.5067}\\
        {0.5318}&{0.5067}&0
        \end{array}} \right]
  \end{equation*}
\end{example}

\subsubsection{Community Large Scale Relationship} 

There are several methods to measure the relationship between different each pair of communities, here the gravity model (GM) \cite{Li2019Identifying} is chosen as the algorithm to obtain the relationship. Based on the obtained AD between each pair of communities, the relationships between communities will be obtained by GM in this part. 

\begin{definition} \label{Def_gravity_model}
  Generally speaking, the community with a large number of nodes would have lower vulnerability. Thus, the gravity index (GI) of each community is obtained by GM, and the details are shown below:
  \begin{equation} \label{Eq_gravity_model}
    {\gamma _i} = \sum\limits_{{c_j} \in C,{c_j} \ne {c_i}} {\frac{{\left| {{N_{{c_i}}}} \right| \times \left| {{N_{{c_j}}}} \right|}}{{\upsilon _{ij}^2}}} 
  \end{equation}
  where ${c_j}$ is all communities in the social network except community ${c_i}$, ${\left| {{N_{{c_i}}}} \right|}$ and ${\left| {{N_{{c_j}}}} \right|}$ are the number of nodes in community ${c_i}$ and ${c_j}$ respectively,  ${\upsilon _{ij}}$ is the AD between ${c_i}$ and ${c_j}$ which is obtained by JSD and LST.
\end{definition}

\begin{example}
  In order to illustrate how GM describes the relationship between communities, one example is given below. With the community division in Example \ref{Example_Q}, the gravity index of community ${c_1}$ can be obtained as follows:
  \begin{equation*}
    {\gamma _1} = \frac{{2 \times 3}}{{{{0.5126}^2}}} + \frac{{2 \times 4}}{{{{0.5318}^2}}} = 51.1223
  \end{equation*}
  and the result about GI for all communities are $\gamma  = [51.1223,69.5717,72.0215]$.
\end{example}

\subsection{Part IV: Vulnerability Evaluating and Ranking} \label{Sub_Method_vulnerability}

\subsubsection{Community Vulnerability Evaluating}

With the obtained EIC, EOC, and GI, the vulnerability of communities can be evaluated and ranked to analyze their property. Because EIC considers the interior information of each community, EOC considers the interaction with neighboring communities which belongs to small scale, and GI considers the relationship between the chosen community and all other communities which is a large scale for information consideration. Thus, the adequate considerations of information enable better measurement of community vulnerability.

\begin{definition} \label{Def_vulnerability}
  The proposed community vulnerability measurement methodology is denoted as ${\zeta _i}$, and defined as follows:
  \begin{equation} \label{Eq_vulnerability}
    {\zeta _i}{\rm{ = }}\frac{1}{{{{\left( {{\eta _i}} \right)}^\alpha }{{\left( {{\sigma _i}} \right)}^\beta }{{\left( {{\gamma _i}} \right)}^\chi }}}
  \end{equation}
  where ${\eta _i}$, ${\sigma _i}$, and ${\gamma _i}$ are normalized EIC, EOC, and GI respectively, which considers the information of communities adequately. The reason for normalizing these factors is to consider them at the same scale. In addition, $\alpha $, $\beta $, and $\chi $ are the weighting parameters for different factors, i.e. the different information considerations. 
  
  The relative vulnerability ${\xi _i}$ of community based on ${\zeta _i}$ is defined as follows:
  \begin{equation} \label{Eq_relative_vulnerability}
    {\xi _i} = \frac{{{\zeta _i}}}{{\mathop {\min }\limits_t {\zeta _t}}}
  \end{equation}
  Thus the relative vulnerability of each community can be shown more detail.
\end{definition}

The setting of these weighting parameters is to evaluate the vulnerability in different situations, which can easily adjust the weight for different information. These weighting parameters make GBCVE more reasonable, and some special cases for ${\zeta _i}$ are shown below:
\begin{enumerate}[1)]
  \item \textbf{$\alpha  = \beta  = \chi $}: ${\zeta _i}$ would consider all factors equally.
  \item \textbf{$\beta  = 1, \alpha  =  \chi  = 0 $}: ${\zeta _i}$ would degenerate classical community vulnerability method ${V_i}$ \cite{Rocco2011Vulnerability}. In addition, ${\zeta _i}$ only considers the interaction with neighboring community in this situation.
  \item \textbf{$\chi  = 0 $}: ${\zeta _i}$ would consider the information in small scale, i.e. the interior information of community and the interaction with neighboring community.
  \item \textbf{$\alpha  = \beta  = 0 $}: ${\zeta _i}$ would consider the information in large scale, i.e. the relationship between chosen community and all other communities.
\end{enumerate}

\begin{example} \label{Example_vulnerability}
  To evaluate the vulnerability of each community in Fig. \ref{Fig_example_network}, this proposed method ${\zeta _i}$ and classical measure ${V_i}$ are used in this example. The weighting factors $\alpha $, $\beta $, and $\chi $ equal to 1 in ${\zeta _i}$ to consider these information equally. According to Eq. (\ref{Eq_vulnerability}) and (\ref{Eq_relative_vulnerability}), the related information, the vulnerability ${\zeta _i}$ and relative vulnerability ${\xi _i}$ of each community of the GBCVE model, and the result of classical measurement ${V_x}$ (vulnerability) and ${R_x}$ (relative vulnerability) are shown in Table \ref{table_vulnerability_example} as follows:

  \begin{table}[!htbp]
    \centering
    \caption{\textbf{The vulnerability of communities in the example social network shown in Fig. \ref{Fig_example_network}.}}
    \resizebox{0.5\textwidth}{!}{
    \begin{tabular}{cccccccc}
    \hline
    \hline
    Community ${c_i}$ & ${\eta _i}$ & ${\sigma _i}$ & ${\gamma _i}$ & ${\zeta _i}$ & ${\xi _i}$ & ${V_i}$ \cite{Rocco2011Vulnerability} & ${R_i}$ \cite{Rocco2011Vulnerability}\\
    \hline
    ${c_1}$           & 0.1667 & 0.5000 & 0.6814 & 17.6099 & 17.6099 & 2 & 2\\
    ${c_2}$           & 0.5000 & 0.5000 & 0.9217 & 4.3133  & 4.3133  & 2 & 2\\
    ${c_3}$           & 1      & 1      & 1      & 1       & 1       & 1 & 1\\
    \hline
    \hline
    \end{tabular}}
    \label{table_vulnerability_example}
  \end{table}

  From the result of ${V_i}$, the classical measure ${V_i}$ can explain ${c_3}$ is the most stable, but it cannot distinguish the vulnerability of ${c_1}$ and ${c_2}$. Thus, ${c_1}$ and ${c_2}$ are equally vulnerable (${V_1} = {V_2}$) for ${V_i}$ perspective. However, as can be seen from Fig. \ref{Fig_example_network}, ${c_1}$ is more vulnerable than ${c_2}$, because there are more nodes in ${c_2}$, and ${c_2}$ is a fully-connected community. 

  From the result of ${\zeta _i}$, this proposed method ${\zeta _i}$ can clearly distinguish the vulnerability of each community. There is a clear order between their vulnerabilities (${\zeta _1} > {\zeta _2} > {\zeta _3}$). Thus, this means this proposed methodology is a more reasonable and effective method to evaluate the vulnerability of each community.
\end{example}

\subsubsection{Community Vulnerability Fuzzy Ranking}

To better compare the property of each community using vulnerability, it is possible to calculate the fuzzy ranking of each community. 

\begin{definition} \label{Def_fuzzy_ranking}
  The fuzzy ranking order for community vulnerability is shown below:
  \begin{equation} \label{Eq_order_FR}
    [{c _{f(1)}}{\rm{ }}{\Omega _1}{\rm{ }}{c _{f(2)}} \cdots {c _{f(\left| C \right| - 1)}}{\rm{ }}{\Omega _{\left| C \right| - 1}}{\rm{ }}{c _{f(\left| C \right|)}}]
  \end{equation}
  where $f(\cdot)$: $\{ 1,2, \cdots ,\left| C \right| - 1\}  \to \{ 1,2, \cdots ,\left| C \right| - 1\} $ is a permutation function that ranks community relative vulnerabilities in the ascending order, i.e. ${\xi _{f(t + 1)}} \ge {\xi _{f(t)}}$ for any $t \in \{ 1,2, \cdots ,\left| C \right| - 1\} $. ${\Omega _1}, \cdots ,{\Omega _{\left| C \right| - 1}}$ is the corresponding relationship between communities, which is the element of $\{  \approx , \le , < , \ll \}$. Inspired by \cite{Capuano2018Fuzzy}, the corresponding relationship transformation approach is defined below:
  \begin{equation} \label{Eq_fuzzy_ranking}
    {\Omega _t} = \left\{ {\begin{array}{*{20}{c}}
      \approx &{{\Delta _t} < 0.25\delta }\\
      \le &{0.25\delta  < {\Delta _t} < 0.75\delta }\\
      < &{0.75\delta  < {\Delta _t} < 1.5\delta }\\
      \ll &{1.5\delta  < {\Delta _t}}
     \end{array}} \right.
  \end{equation}
  where ${{\Delta _t}}$ is the distinction between adjacent vulnerability, i.e. ${\Delta _t} = {\xi _{f(t + 1)}} - {\xi _{f(t)}}$. In addition, the $\delta $ is determined by the average of distinction between adjacent vulnerability, which is shown as follows,
  \begin{equation} \label{Eq_delta}
    \delta  = \frac{1}{{\left| C \right| - 1}}\sum\limits_{t = 1}^{\left| C \right| - 1} {{\Delta _t}}  = \frac{1}{{\left| C \right| - 1}}\sum\limits_{t = 1}^{\left| C \right| - 1} {\left( {{\xi _{f(t + 1)}} - {\xi _{f(t)}}} \right)}
  \end{equation}
  Thus the fuzzy ranking order of community vulnerability can be obtained.
\end{definition}

\begin{example}
  The ${\xi_i}$ in Example \ref{Example_vulnerability} is used in this example to show the fuzzy ranking of the example social network. The  $\delta $ can be obtained below:
  \begin{equation*}
    \begin{split}
      \displaystyle \delta  & = \frac{1}{{3 - 1}}\left( {\left( {4.3113 - 1} \right) + \left( {17.6099 - 4.3113} \right)} \right)\\
      \displaystyle & = 8.3050
      \end{split}
  \end{equation*}
  Then, the corresponding relationship is determined as follows,
  \begin{equation*}
    \begin{split}
      \displaystyle {\Delta _1} & = 4.3113 - 1 = 3.3113 \\
      \displaystyle {\Delta _2} & = 17.6099 - 4.3113 = 13.2986
      \end{split}
  \end{equation*}
  Based on Eq. (\ref{Eq_fuzzy_ranking}), the ${\Omega _1}$ and ${\Omega _2}$ would be $ \le $ and $ \ll $ respectively, and the fuzzy ranking order of three communities is $[{c_3} \le {c_2} \ll {c_1}]$. From the fuzzy ranking order, the vulnerability of ${c_3}$ and ${c_2}$ are similar, but the vulnerability of ${c_1}$ is far greater than the other two VoC.
\end{example}

\subsection{Part V: sensitivity Analysis} \label{Sub_Method_sensitive}

After the VoCs are obtained by GBCVE, the sensitivity of this proposed method should be analyzed. Because these three weighting parameters can change the result of VoCs with different values, they are significant for VoCs. Thus, how to determine these weighting parameters is still an important issue for this model. In mathematical and physical models, the input influences on the output variance can be analyzed by global sensitivity analysis. In this paper, the \emph{Sobol' indices} (SI) based on variance decomposition is used. 

\begin{definition} \label{Def_Sobol}
  The \emph{first-order Sobol' index} ${S_i}$ and \emph{total effect index} ${S_{{T_i}}}$ are detailed introduced below:
  \begin{equation} \label{Eq_Sobol}
    \begin{split}
      \displaystyle {S_i} & = \frac{{\mathop {Var}\nolimits_{{X_i}} \left( {\mathop E\nolimits_{{X_{ \sim i}}} \left( {Y|{X_i}} \right)} \right)}}{{Var\left( Y \right)}}\\
      \displaystyle {S_{{T_i}}} & = \frac{{\mathop E\nolimits_{{X_{ \sim i}}} \left( {\mathop {Var}\nolimits_{{X_i}} \left( {Y|{X_{ \sim i}}} \right)} \right)}}{{Var\left( Y \right)}}
      \end{split}
  \end{equation}
  where ${{X_i}}$ is the \emph{i}th independent factor of input $X$, ${{X_{ \sim i}}}$ is all inputs except ${{X_i}}$, $Y$ is the output of this model, and ${Var\left( Y \right)}$ is the variance of $Y$ with the change of input. Specifically, ${S_i}$ can describe the contribution of ${{X_i}}$ to $Y$. ${S_{{T_i}}}$ can describe the contribution of the variability of ${{X_i}}$ to the variance of $Y$, i.e. ${Var\left( Y \right)}$. In addition, ${S_{{T_i}}}$ considers both individual input effects and the interaction with other inputs.
\end{definition}

\section{Evaluations}\label{Sec_evaluations}

In this section, two real world complex networks are used to show the effectiveness and reasonability of this proposed model, and they are Manzi network and Italian power network respectively. Both of them are commonly used networks for analyzing vulnerabilities, and they are more complex than the previous example network.






\subsection{Case I: Manzi Network}

\begin{figure}[!htbp]
  \centering
  \includegraphics[width=0.5\textwidth]{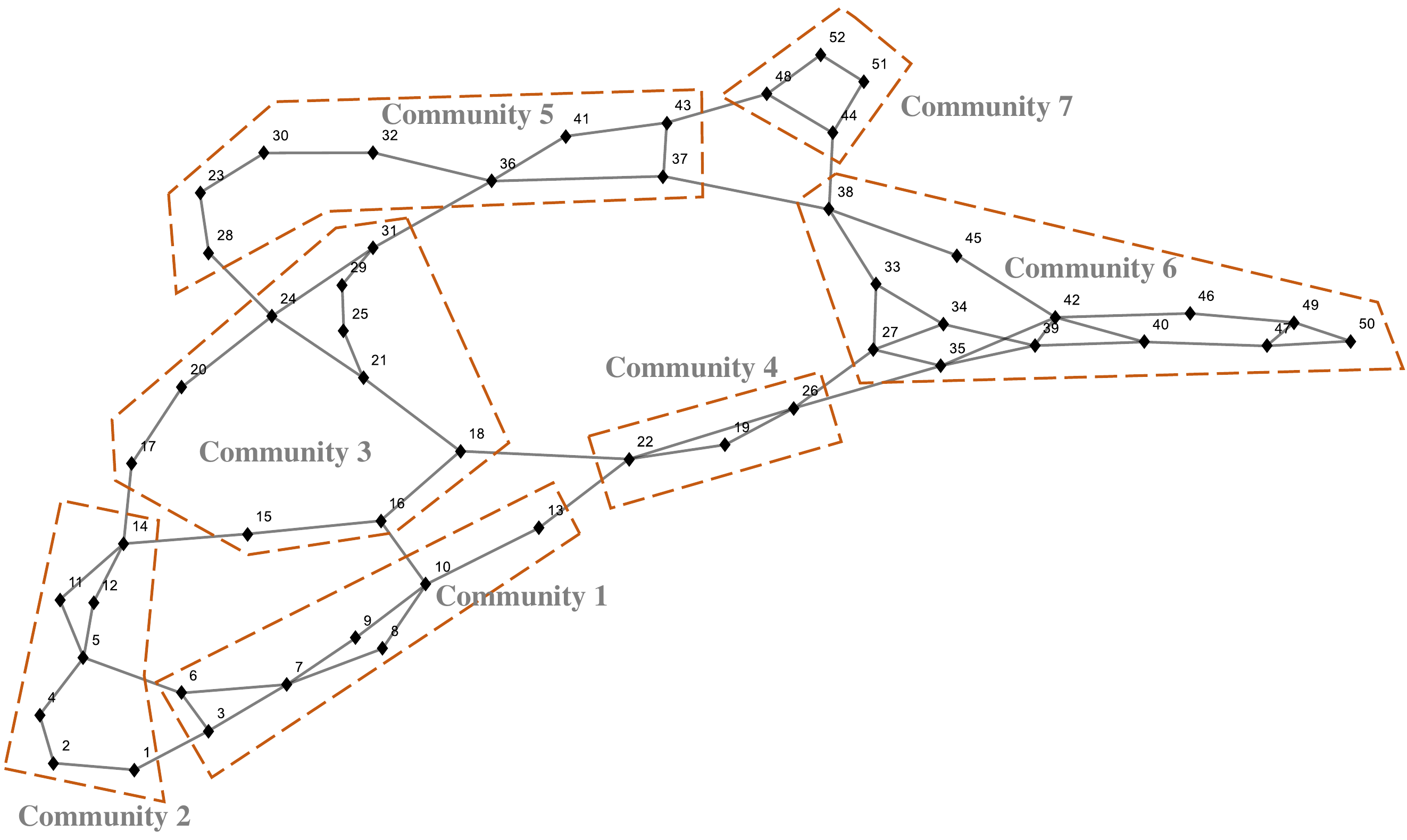}\\
  \caption{\textbf{The structure and community division of Manzi network.}}
  \label{Fig_Manzi}
\end{figure}

The telephone network in Belgium \cite{Manzi2001Fishman} is used in this subsection to analyze these communities' vulnerability. The Manzi network structure is shown in Fig. \ref{Fig_Manzi}, and there are lots of nodes and edges in this network, which is more complex than the example social network. The purpose of this model is to evaluate and rank the vulnerability of each community.

The Manzi network is divided into several communities by Newman modularity, and the detail community division is shown in Fig. \ref{Fig_Manzi}. Observed from Fig. \ref{Fig_Manzi}, this network is divided into seven communities, and the modularity $Q$ equals to 0.6316, which means the community structure is strong. The specific nodes in each community can be also observed in Fig. \ref{Fig_Manzi}.

We first calculate the AD between each pair of communities rather than the basic information of the community, because the basic information can be easily obtained by the topological structure in Fig. \ref{Fig_Manzi}, which will be given directly in Table \ref{table_manzi_result}. According the definition as Eq. (\ref{Eq_JSD}), it is clear to get the JSD between communities, and the AD after LST for each pair of communities can be obtained by Eq. (\ref{Eq_log}) with $\varphi = 3$. The AD describes the difference between communities, and it is symmetrical which can describe the distance accurately. The details of $\upsilon $ can be shown below:

  \begin{equation*}
    \resizebox{0.5\textwidth}{!}{
     $\upsilon {\rm{ = }}\left[ {\begin{array}{*{20}{c}}
      {0.0000}&{0.5027}&{0.5092}&{0.5284}&{0.5080}&{0.5196}&{0.5176}\\
      {0.5027}&{0.0000}&{0.5088}&{0.5342}&{0.5042}&{0.5189}&{0.5172}\\ 
      {0.5092}&{0.5088}&{0.0000}&{0.5514}&{0.5055}&{0.5057}&{0.5343}\\ 
      {0.5284}&{0.5342}&{0.5514}&{0.0000}&{0.5445}&{0.5715}&{0.5067}\\ 
      {0.5080}&{0.5042}&{0.5055}&{0.5445}&{0.0000}&{0.5124}&{0.5255}\\ 
      {0.5196}&{0.5189}&{0.5057}&{0.5715}&{0.5124}&{0.0000}&{0.5532}\\ 
      {0.5176}&{0.5172}&{0.5343}&{0.5067}&{0.5255}&{0.5532}&{0.0000} 
      \end{array}} \right]$
      }
  \end{equation*}

After the AD between communities is obtained, these communities will construct a CN, and this CN with seven communities for Manzi network can be shown in Fig. \ref{Fig_community_network} as follows:


\begin{figure}[!htbp]
  \centering
  \includegraphics[width=0.5\textwidth]{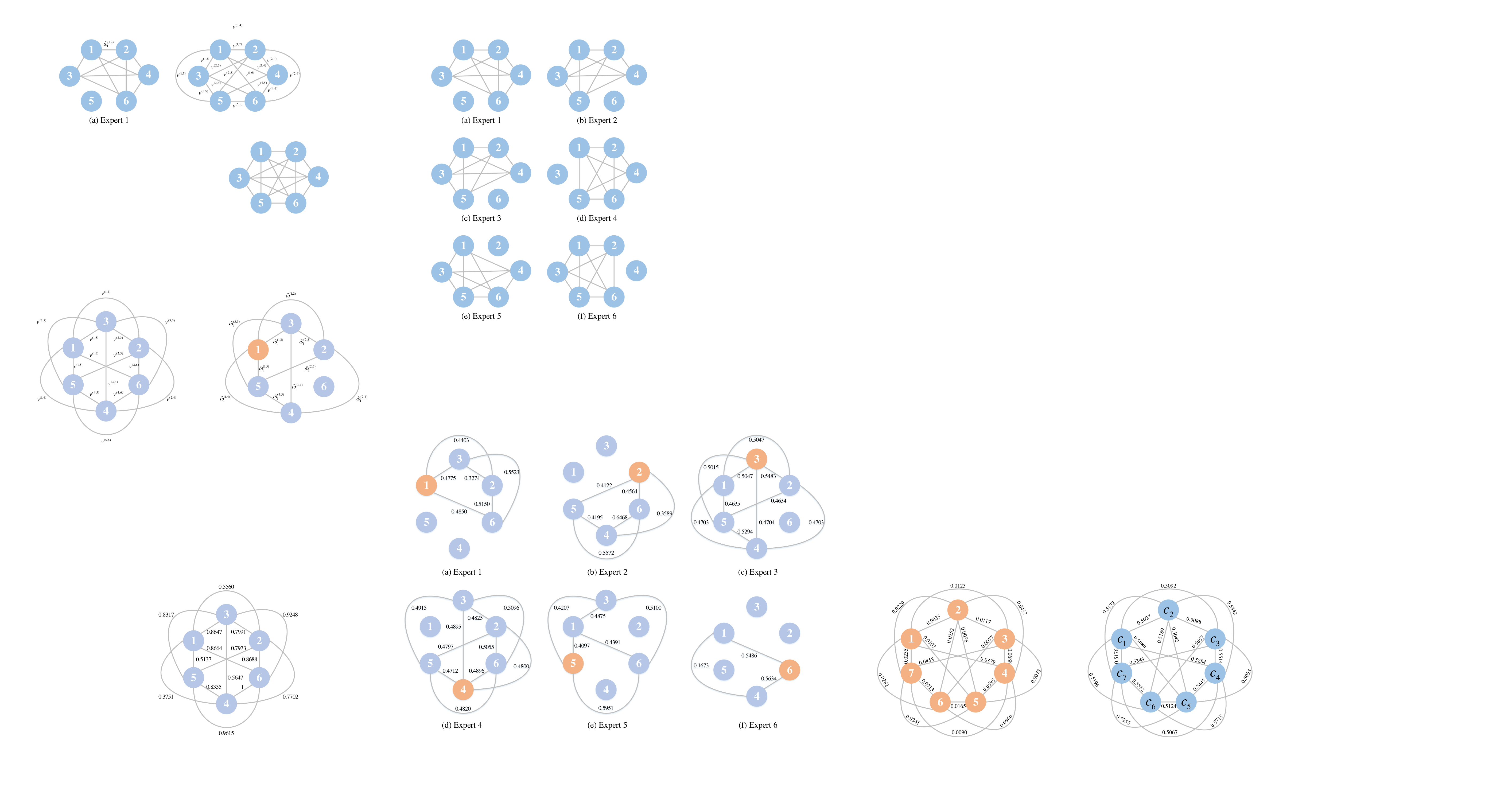}\\
  \caption{\textbf{The community network of Manzi network.}}
  \label{Fig_community_network}
\end{figure}

It can be found that the CN is a fully-connected undirected network, this is because the AD considers the relationship between each pair of community in the network. The relationship in the CN is the existence of the edge.

Then, with the obtained AD and the number of nodes in each community, the GI defined in Eq. (\ref{Eq_gravity_model}) of each community can be obtained as follows:

\begin{equation*}
  \displaystyle \gamma {\rm{ = }}\left[ {\begin{array}{*{20}{c}}
    {1.1976 \times {{10}^3}}\\
    {1.2009 \times {{10}^3}}\\
    {1.6005 \times {{10}^3}}\\
    {4.9455 \times {{10}^2}}\\
    {1.3434 \times {{10}^3}}\\
    {1.8688 \times {{10}^3}}\\
    {6.8179 \times {{10}^2}}
    \end{array}} \right]
\end{equation*}

The EIC and EOC can be obtained directly from the adjacency matrix of the network, and GI has been just obtained. These factors are normalized first, then the vulnerability ${\zeta _i}$ and relative vulnerability ${\xi _i}$ can be obtained by the GBCVE model through Eq. (\ref{Eq_vulnerability}) and Eq. (\ref{Eq_relative_vulnerability}). It is worth noticing that three weighting parameters $\alpha ,\beta ,\chi $ are set to 1 here, which is an equal consideration for all factors. The information of communities, obtained ${\zeta _i}$ and ${\xi _i}$, the result of classical measurement ${V_x}$ and ${R_x}$, and the ranking result of this proposed GBCVE model are shown in Table \ref{table_manzi_result}. 

\begin{table}[!htbp]
  \centering
  \caption{\textbf{The related information and vulnerability results of seven community in Manzi network.}}
  \resizebox{0.5\textwidth}{!}{
  \begin{tabular}{ccccccccc}
  \hline
  \hline
  Community  & ${\eta _i}$ & ${\sigma _i}$ & ${\gamma _i}$ & ${\zeta _i}$ & ${\xi _i}$ & ${V_i}$ \cite{Rocco2011Vulnerability} & ${R_i}$ \cite{Rocco2011Vulnerability} & Ranking \\
  \hline
  ${c_1}$  & 0.4211 & 0.6667 & 0.6409 & 5.5590 & 3.7060 & 1.5000 & 1.5000  & 4 \\
  ${c_2}$  & 0.3684 & 0.6667 & 0.6426 & 6.3358 & 4.2239 & 1.5000 & 1.5000  & 3 \\
  ${c_3}$  & 0.5263 & 1.0000 & 0.8565 & 2.2184 & 1.4790 & 1.0000 & 1.0000  & 6 \\
  ${c_4}$  & 0.1579 & 0.6667 & 0.2646 & 35.8975 & 23.9316 & 1.5000 & 1.5000  & 2 \\
  ${c_5}$  & 0.4211 & 0.6667 & 0.7188 & 4.9559 & 3.3039 & 1.5000 & 1.5000  & 5 \\
  ${c_6}$  & 1.0000 & 0.6667 & 1.0000 & 1.5000 & 1.0000 & 1.5000 & 1.5000 & 7 \\
  ${c_7}$  & 0.2105 & 0.3333 & 0.3648 & 39.0586 & 26.0391 & 3.0000 & 3.0000  & 1 \\
  \hline
  \hline
  \end{tabular}}
  \label{table_manzi_result}
\end{table}

Observed from Table \ref{table_manzi_result}, the relative vulnerability of each community is ${\xi _1} = 3.7060,{\xi _2} = 4.2239,{\xi _3} = 1.4790,{\xi _4} = 23.9316,{\xi _5} = 3.3039,{\xi _6} = 1,{\xi _7} = 26.0391$. Based on Eq. (\ref{Eq_delta}), $\delta$ can be calculated as 4.1732. Finally, the community vulnerability fuzzy ranking order can be obtained by Eq. (\ref{Eq_fuzzy_ranking}) as: $ {c_6}  \approx  {c_3}  \le  {c_5} \approx  {c_1}  \approx  {c_2}  \ll  {c_4} \le  {c_7} $.

From ${\xi _i}$ and ${R_i}$, it can be found that ${c_7}$ is the most vulnerable community recognized by two methods, because there is only a ring of four nodes. Then ${R_i}$ believes that ${c_1}$, ${c_2}$, ${c_4}$, ${c_5}$, and ${c_6}$ has the same vulnerability, but it is impossible because of their different structure. This proposed method ${\xi _i}$ can give a specific ranking of these communities' vulnerability. ${c_4}$ is the second most vulnerable (fully connected network with three nodes). In addition, ${VoC_7}$ and ${VoC_4}$ are far greater than other communities' vulnerability from the fuzzy ranking result. The vulnerability of ${c_1}$, ${c_2}$, and ${c_5}$ are close (similar with ${R_i}$), but there is a specific ranking, i.e. ${\xi _5} < {\xi _1} < {\xi _2}$, which is better than the classical method ${R_i}$. The ${VoC_1}$, ${VoC_2}$, and ${VoC_5}$ can be observed from the structure of each community from Fig. \ref{Fig_Manzi}. Another difference between ${\xi _i}$ and ${R_i}$ is the result about ${c_6}$ and ${c_3}$. ${R_i}$ thinks ${c_6}$ is more vulnerable than ${c_3}$, and this proposed method ${\xi _i}$ thinks ${c_3}$ is more vulnerable than ${c_6}$ and their vulnerability is similar from the fuzzy ranking. Observed from Fig. \ref{Fig_Manzi}, there is only a ring in ${c_3}$ and other nodes are connected in a straight line, and ${c_3}$ will have a high probability of being split into two subgraphs after removing a node; most of the nodes in ${c_6}$ are connected in the form of a ring, and it will rarely be divided into two subgraphs after removing a node. Thus, it can be clearly seen that ${c_3}$ is more vulnerable than ${c_6}$ from the structure of these communities. Therefore, from the comparison of these two methods in Manzi network, ${\xi _i}$ can give a specific ranking result of each community, and it can show how big the gap between $VoCs$ is from the fuzzy ranking. Thus this proposed model is a more reasonable and effective method to evaluate the vulnerability of each community in the social network.



\begin{table}[!htbp]
  \centering
  \caption{\textbf{The sensitivity analysis results of the $VoCs$ with weighting parameters $\alpha $, $\beta $, and $\chi $.}}
  \resizebox{0.5\textwidth}{!}{
  \begin{tabular}{ccccccc}
  \hline
  \hline
  Community & ${S_i}(\alpha )$ & ${S_{{T_i}}}(\alpha )$ & ${S_i}(\beta )$ & ${S_{{T_i}}}(\beta )$ & ${S_i}(\chi )$ & ${S_{{T_i}}}(\chi )$ \\
  \hline
  ${c_1}$ &  0.4166 & 0.7287 & 0.1078 & 0.3099 & 0.1287 & 0.3669  \\
  ${c_2}$ &  0.4431 & 0.7733 & 0.0911 & 0.2971 & 0.1078 & 0.3491  \\
  ${c_3}$ &  0.9004 & 0.9410 & 0.0000 & 0.0000 & 0.0598 & 0.1007  \\
  ${c_4}$ &  0.2035 & 0.8325 & 0.0156 & 0.2355 & 0.1272 & 0.7504  \\
  ${c_5}$ &  0.4907 & 0.7632 & 0.1274 & 0.3254 & 0.0863 & 0.2463  \\
  ${c_6}$ &  0.0000 & 0.0000 & 0.9991 & 0.9990 & 0.0000 & 0.0000  \\
  ${c_7}$ &  0.1192 & 0.7722 & 0.0688 & 0.6419 & 0.0585 & 0.6374  \\
  \hline
  \hline
  \end{tabular}}
  \label{table_manzi_sensitive}
\end{table}

Then, SI in Section \ref{Sub_Method_sensitive} is used to analyze the sensitivity of these weighting parameters. Different values of weighting parameters can focus on different aspects of information of community, and give different vulnerability results for networks. The sensitivity analysis results of VoCs with three weighting parameters are shown in Table \ref{table_manzi_sensitive}, and observed from the results, some conclusions can be obtained below:

\begin{enumerate}[1)]
  \item The value of ${S_i}$ can show the sensitivity of each weighting parameter. For example, ${VoC_1}$ is most sensitive to $\alpha$, followed by $\beta$, and $\chi$ is the least sensitive.
  \item When the value of factor equals to 1, the ${S_i}$ and ${S_{{T_i}}}$ of weighting parameter would equal to 0, like ${\eta _6} = 1$ cause ${S_6}(\alpha ) = 0$ ${S_{{T_6}}}(\alpha ) = 0$. This means that when the factor equals to 1, the sensitivity of weighting parameters (index) would not change and equal to 0.
  \item ${S_i}(\alpha )$ is larger than ${S_i}(\beta )$ and ${S_i}(\chi )$, which means this method is more sensitive with $\alpha$, i.e. EIC is more influential to VoCs.
  \item In most cases, the sum of ${S_i}$ over three weighting parameters in each community is less than 1. This situation is due to the interaction between these consideration parameters, but this situation does not appear in ${S_{{T_i}}}$.
  \item Even if there is an interaction that makes ${S_i}$ and ${S_{{T_i}}}$ different, it is interesting that ${S_i}$ and ${S_{{T_i}}}$ maintain the same order between different information factors in the same community. For example, in ${c_1}$, ${S_1}(\alpha ) > {S_1}(\chi ) > {S_1}(\beta )$ and ${S_{{T_1}}}(\alpha ) > {S_{{T_1}}}(\beta ) > {S_{{T_1}}}(\chi )$.
\end{enumerate}


\subsection{Case II: Italian Network}

\begin{figure}[!htbp]
  \centering
  \includegraphics[width=0.5\textwidth]{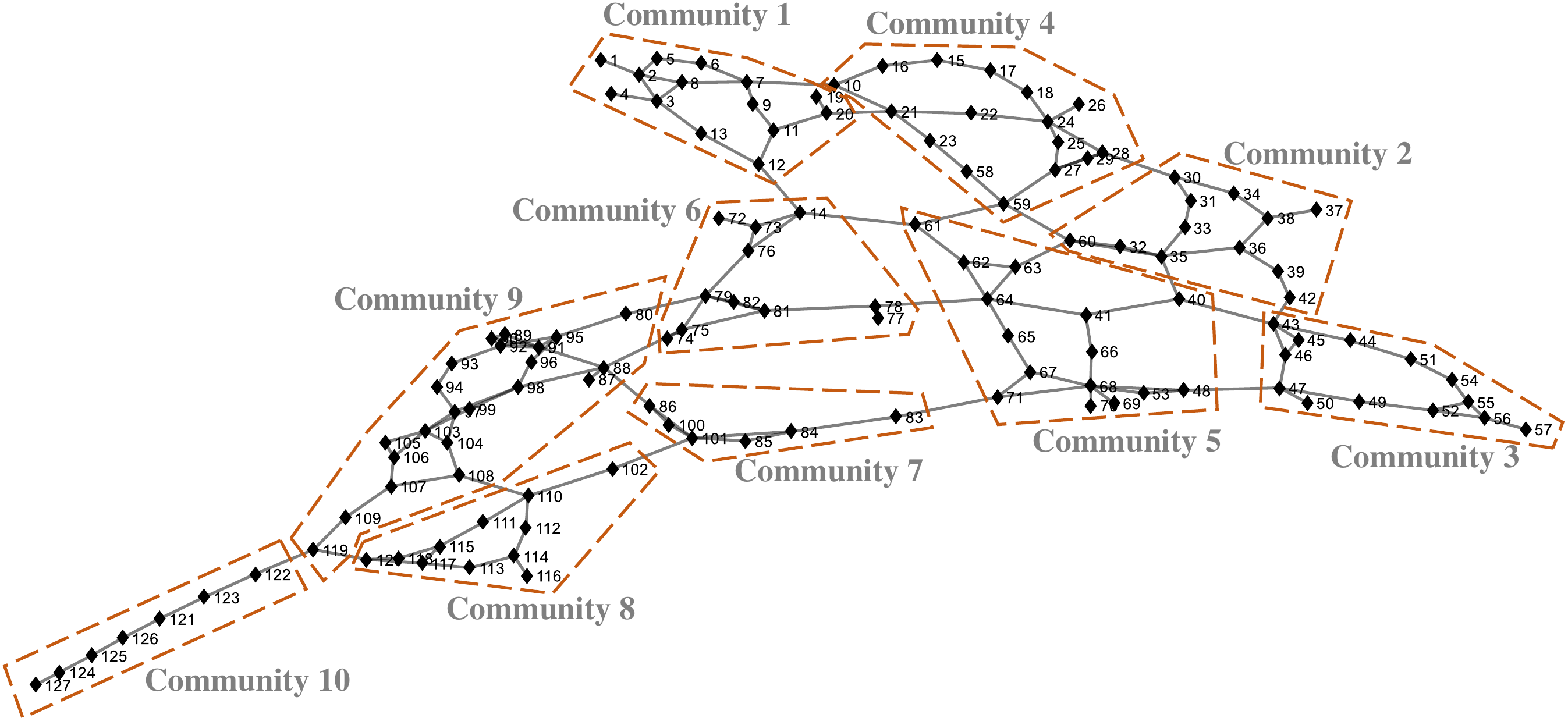}\\
  \caption{\textbf{The structure and community division of Italian network.}}
  \label{Fig_Italian}
\end{figure}

Then, Wei \cite{Wei2018Measuring} method is also used in this subsection to prove the superiority of this proposed GBCVE model. The Italian 380KV power transmission grid network \cite{Crucitti2005Locating}, a network often used for vulnerability analyzing, is used in this subsection as the subject of the experiment. The structure and the community division of Italian network are shown in Fig. \ref{Fig_Italian}. This network is divided into 10 communities by Newman's modularity method and the modularity $Q$ equals to 0.7596 which shows the strong community structure of Italian network. The specific nodes of each community can be seen in Fig. \ref{Fig_Italian}. Similarly, $\varphi$ is set to 3 in the LST function to obtain AD between communities, and $\alpha ,\beta ,\chi $ are set to 1, which means the same consideration of information.

According to Section \ref{Sec_methodology}, the related information and the vulnerability of each community obtained by different methods are shown in Table \ref{table_Italian_result}. Furthermore, the fuzzy ranking result of Italian network is ${c_9} \approx  {c_5} \approx  {c_4} \approx  {c_2} \approx  {c_6} \approx  {c_1} \approx  {c_3} \le {c_8} \ll {c_7} \ll {c_{10}}$.

\begin{table}[!htbp]
  \centering
  \caption{\textbf{The related information and vulnerability results of ten community in Italian network.}}
  \resizebox{0.5\textwidth}{!}{
  \begin{tabular}{ccccccccccc}
  \hline
  \hline
  Community  & ${\eta _i}$ & ${\sigma _i}$ & ${\gamma _i}$ & ${\zeta _i}$ & ${\xi _i}$ & ${V_i}$ \cite{Rocco2011Vulnerability} & ${R_i}$ \cite{Rocco2011Vulnerability} & ${V_i}^\prime $ \cite{Wei2018Measuring} & ${R_i}^\prime $ \cite{Wei2018Measuring} &Ranking \\
  \hline
  ${c_1}$  & 0.5161 & 0.3750 & 0.7061 & 7.3168 & 5.4876 & 2.6667 & 2.6667 & 9.1020 & 2.6547 & 5 \\
  ${c_2}$  & 0.4194 & 0.6250 & 0.6117 & 6.2370 & 4.6778 & 1.6000 & 1.6000 & 5.0539 & 1.4741 & 7 \\
  ${c_3}$  & 0.4839 & 0.3750 & 0.6551 & 8.4123 & 6.3092 & 2.6667 & 2.6667 & 6.9333 & 2.0222 & 4 \\
  ${c_4}$  & 0.6129 & 0.6250 & 0.7792 & 3.3502 & 2.5126 & 1.6000 & 1.6000 & 5.6606 & 1.6510 & 8 \\
  ${c_5}$  & 0.5806 & 1.0000 & 0.7404 & 2.3262 & 1.7447 & 1.0000 & 1.0000 & 4.0843 & 1.1912 & 9 \\
  ${c_6}$  & 0.4194 & 0.6250 & 0.5617 & 6.7920 & 5.0940 & 1.6000 & 1.6000 & 3.4459 & 1.0050 & 6 \\
  ${c_7}$  & 0.2258 & 0.3750 & 0.3006 & 39.2892 & 29.4669 & 2.6667 & 2.6667 & 3.4286 & 1.0000 & 2 \\
  ${c_8}$  & 0.3871 & 0.3750 & 0.5634 & 12.2266 & 9.1699 & 2.6667 & 2.6667 & 8.5554 & 2.4953 & 3 \\
  ${c_9}$  & 1.0000 & 0.7500 & 1.0000 & 1.3333 & 1.0000 & 1.3333 & 1.3333 & 9.5387 & 2.7821 & 10 \\
  ${c_{10}}$  & 0.1935 & 0.1250 & 0.3567 & 115.8782 & 86.9087 & 8.0000 & 8.0000 & 39.2052 & 11.4347 & 1 \\
  \hline
  \hline
  \end{tabular}}
  \label{table_Italian_result}
\end{table}

It can be found that ${c_{10}}$ is the most vulnerable community in the network evaluated by three methods, which is the common point of the three methods. Then, in order to compare the performance of vulnerability rankings obtained by different methods, the detailed ranking results are shown in Table \ref{table_Italian_ranking}. 

\begin{table}[!htbp]
  \centering
  \caption{\textbf{The vulnerability ranking of communities obtained by different methods in Italian network.}}
  \resizebox{0.5\textwidth}{!}{
  \begin{tabular}{cc}
  \hline
  \hline
  Method  & Vulnerability Ranking \\
  \hline
  Classical method \cite{Rocco2011Vulnerability} & ${R_5} < {R_9} < {R_2} = {R_4} = {R_6} < {R_1} = {R_3} = {R_7} = {R_8} < {R_{10}}$  \\
  Wei \emph{et al.} method \cite{Wei2018Measuring}  & ${R_7}^\prime < {R_6}^\prime < {R_5}^\prime < {R_2}^\prime < {R_4}^\prime < {R_3}^\prime < {R_8}^\prime < {R_1}^\prime < {R_9}^\prime < {R_{10}}^\prime $ \\
  Proposed method & ${\xi _9} < {\xi _5} < {\xi _4} < {\xi _2} < {\xi _6} < {\xi _1} < {\xi _3} < {\xi _8} < {\xi _7} < {\xi _{10}} $ \\
  Fuzzy ranking of proposed method  & ${c_9} \approx  {c_5} \approx  {c_4} \approx  {c_2} \approx  {c_6} \approx  {c_1} \approx  {c_3} \le {c_8} \ll {c_7} \ll {c_{10}}$ \\
  \hline
  \hline
  \end{tabular}}
  \label{table_Italian_ranking}
\end{table}

Observed from Table \ref{table_Italian_ranking}, compared with the classical method and this proposed method, Wei method gives a more confusing result. Specifically, ${c_9}$ is considered to be a less vulnerable community by ${R_i}$ (Rank 9) and ${\xi _i}$ (Rank 10), but in Wei method it is considered to be a highly vulnerable community (Rank 2). In addition, ${c_7}$ is a more vulnerable community in ${R_i}$ (Rank 3) and ${\xi _i}$ (Rank 2), but a different judgment in ${R_{i}}^\prime$ (Rank 10). Therefore, Wei method is a relatively confusing method, and the rest of ranking in ${R_{i}}^\prime$ would not be analyzed. 

Compared with classical method, this proposed method can describe the change of VoCs more detail. For instance, ${R_i}$ gives equal vulnerability of ${c_1}$, ${c_3}$, ${c_7}$, ${c_8}$, and cannot distinguish the difference of their vulnerability, but this proposed method can give a specific ranking of their vulnerability, i.e. ${\xi _1} < {\xi _3} < {\xi _8} < {\xi _7}$, and their fuzzy ranking ${c_1} \approx  {c_3} \le {c_8} \ll {c_7}$. It can be found that this proposed method not only gives a change in the vulnerability between these communities, but also a specific description of the relationship between them, i.e. $\ll$ or $\le$ or $\approx$. This is also suitable for other communities whose vulnerability cannot be evaluated by classical method, i.e. ${c_2}$, ${c_4}$, and ${c_6}$.

Therefore, compared with classical method and Wei method, this proposed GBCVE model can get vulnerability result of communities more effectively. In addition, the result obtained by GBCVE model is more reasonable which follows the rule of structure.

\section{Conclusion}\label{Sec_conclusion}

The vulnerability measuring is definitely essential of community study of social networks. In most cases, the vulnerability result of the community that takes into account more kinds of factors is more credible. In this paper, a novel community vulnerability evaluation model named GBCVE is proposed to consider adequate factors of each community. In this proposed method, these three factors are the number of edges inside the community, the number of edges connected neighboring communities, and the gravity index, which correspond to interior information of the community, small scale interaction relationship, and large scale interaction relationship. The first two factors are the basic property of the network which are easy to obtain from the topological structure, and the last one factor (large scale interaction relationship) is the focus of this proposed GBCVE model. The structure difference between each pair of community is firstly measured by Jensen-Shannon divergence. Then the difference is converted to abstract distance by the log-sigmoid transition function ($\varphi = 3$). The gravity index of each community would be obtained by the gravity model lastly which shows the relationship between the chosen community and all other communities. Furthermore, the vulnerability degree of each community can be evaluated by this proposed GBCVE model (weighting parameters equal to 1 in general), and the specific vulnerability ranking is given via fuzzy ranking technique. In some cases, this proposed method can degenerate to classical evaluation method with special setting of weighting parameters. In addition, the global sensitivity of weighting parameters is analyzed by Sobol' index.

As seen in Section \ref{Sec_evaluations}, two real world complex networks are used to show the effectiveness and reasonability of this proposed model. For Manzi network, there are 7 communities to evaluate their vulnerability. It is obvious that this proposed model can better evaluate the vulnerability of each community, because it can give different values of ${\xi _i}$ when classical method only judges the same vulnerability of these communities. In addition, the fuzzy ranking of communities can show how much the difference of vulnerability between them is. The sensitivity analysis shows which factor is more influential to the vulnerability result. For Italian network, compared with the confusing result in Wei method, the vulnerability result of this proposed model is more reasonable and objective. 

Considering the important position of vulnerability evaluating in the community study of social networks, we believe that this proposed model can achieve more reasonable and objective results in the community research. Specifically, it is meaningfully to use the community vulnerability order with the combination of other models, such as the community recovery and alliance partner looking. In addition, other important factors to community properties in overlapping community structure are also worth exploring, which is the focus of our future research.

\section*{Acknowledgment}

The authors would like to thank XXX.

\ifCLASSOPTIONcaptionsoff
  \newpage
\fi

\bibliographystyle{IEEEtran}
\bibliography{mybibfile}

\begin{IEEEbiographynophoto}{Tao Wen}
Biography text here.
\end{IEEEbiographynophoto}

\end{document}